\documentstyle[aps,manuscript]{revtex}
\draft
\begin{document}

\title{Covariant description for superfluids \\
in gravitational fields}
\author{H. Casini and R. Montemayor \\
Instituto Balseiro and Centro At\'{o}mico Bariloche, \\
Universidad Nacional de Cuyo and CNEA\\
8400 - S.C. de Bariloche\\
R\'{\i}o Negro, Argentina}
\maketitle
\begin{abstract}
In this paper we develop a formalism to describe a superfluid in a
gravitational background. This formalism is based on a covariant
generalization of the field description for a superconductor
in terms of a U(1) spontaneous symmetry breaking. We study the stability of
the solutions for a vortexless fluid and the force acting on vortices in the
fluid, which is a generalization of the well-known flat space-time Magnus
force. To clarify the development we include the explicit discussion of two
particular cases, one of them of astrophysical interest.
\end{abstract}

\section{Introduction}

In this paper we study the behavior of a superfluid in presence of a
gravitational field. Under the usual conditions that can be obtained in a
laboratory only helium shows the superfluid effect, because this is the only
substance that remains liquid at temperatures low enough for quantum effects
to dominate. New and appealing scenarios with very interesting possibilities
are provided by some astrophysical systems. One scenario is realted to
neutron stars, where there are very high pressures and densities, and thus
where superfluid effects can appear at relatively high temperatures. A widely
accepted description for the constituent matter of a neutron star considers
two coexisting quantum superfluids, neutron and proton seas, where quantum
dynamics plays an essential role\cite{pines}. But, in contrast with
laboratories on Earth, there we have not only superfluids but also strong
gravitational fields. This makes it possible to have significant
gravimagnetic effects, which could produce phenomena similar to the ones that
appear in the case of the superconductors\cite{casini}. We can also mention a
more speculative scenario, of interest at a cosmological level, which is the
formation of a condensate in the sea of relic neutrinos\cite {guinzburg}.
Scenarios like these open the possibility of gaining access to a
phenomenology which would allow a deeper understanding of different aspects
of the gravitational interaction in quantum systems.

There have already been some attempts to describe a covariant superfluid,
usually stressing classical geometrical aspects of the
problem\cite{langlois}. We present here a different approach, which is a
covariant generalization of the description of the superconductor
phenomenology based on a U(1) spontaneous symmetry breaking\cite{weinberg1}.
This approach allows us to discuss dynamical aspects of the superfluids in a
gravitational background, on a very general and well established basis, with
a clear phenomenological interpretation.

In the following section we develop the general formalism for a superfluid
in a gravitational background, and discuss the dynamics of the scalar
bosonic excitations. We make explicit the corresponding equation of motion
and the conditions that warrant the stability of the ground state. In
Section 3 we consider the presence of vortices and study the forces acting
on them, using a Kalb-Ramond description for the vortices\cite{davis}. In
particular, we obtain the force due to the background, which is a
generalization of the well known Magnus force, and the stable vortex
configurations. Section 4 contains the analysis of two cases, a superfluid
in a laboratory on Earth and in a neutron star, which exemplify the general
formalism. Finally, in the last section we make some comments and remarks.

\section{Superfluid in a gravitational background}

The superfluidity phenomena is related at the microscopic level to the
formation of a condensate that spontaneously breaks a global symmetry. The
spontaneous breaking of the symmetry leads to a Nambu-Goldstone excitation
with zero energy in the limit of vanishing momentum. We develop here a
presentation for the superfluid that follows the one for superconductors in
flat space-time given in Ref. \cite{weinberg1}. We start with a fluid which
is formed by particles represented by a matter field $\Psi $ with the
simplest U(1) global symmetry group $\Psi (x)\longrightarrow e^{i\Lambda
}\,\Psi (x)$. To be specific we can think of the neutron fluid in a neutron
star and the baryonic symmetry that leads to the baryon number conservation
in the system. We can always perform a field splitting
\begin{equation}
\Psi (x)=e^{i\phi (x)\,\,}\psi (x)\;.  \label{split}
\end{equation}
The group transformation acting on the boson $\phi $ is $\phi
(x)\longrightarrow \phi (x)+\Lambda$.

The U(1) invariant density Lagrangian must be a function of the derivatives
of $\phi $ and the U(1) invariant field $\psi $. The most general Lagrangian
density allowed by the symmetries is a nonlocal function of the field, where
the nonlocality extends over a range of the order of the coherence length of
the superfluid. Given that we are interested in the macroscopic fluid
motion, we will only consider local terms in the Lagrangian that effectively
describe the long range behavior. Such terms must be scalars and should be
constructed as a contraction of covariant quantities. The only possible
fields involved are the gradient of $\phi $, the metric and a number of
tensors $\lambda ^{\mu }$, $\lambda ^{\mu \nu }$, etc., which depend on the
field $\psi $ and must satisfy the additional requirements of symmetry
eventually imposed by this field. Thus, the Lagrangian can be written as the
expansion

\begin{equation}
{\cal L}={\cal L}_{o}[\psi ]+\lambda ^{\mu }[\psi ]\phi _{,\mu }+\frac{1}{2}
\lambda ^{\mu \nu }[\psi ]\,\phi _{,\nu }\phi _{,\mu }+...\,\,\,\,\,\,\,\,,
\label{lag}
\end{equation}
where ${\cal L}_{o}$ is a $\phi $ independent Lagrangian. The symmetry
breaking implies that the Goldstone field $\phi $ has an effective dynamics
below some typical energy where the condensate forms and where the tensor
$\lambda^{\mu\nu}$ develops a non null expectation value. The dynamics of
the superfluid is completely described by the Goldstone boson because the
gap makes $\psi $ nondynamical at low energies. However, the expectation
values of the tensors $\lambda ^{\mu \nu }$ and $\lambda ^{\mu }$ generally
depend on external forces, such as the ones induced by the gravitational
field and the ones on the background state. That is, in general $\lambda
^{\mu \nu }$ and $\lambda ^{\mu }$ should satisfy classical equations that
involve the gravitational field, of the type given by the hydrostatic
equilibrium, while the Goldstone boson should satisfy the full quantum
dynamics given by the Lagrangian (\ref{lag}). Besides this, if the Goldstone
field satisfies a four dimensional dynamics the $\lambda ^{\mu \nu }$ tensor
must be non degenerate. The effective theory makes contact with the
microscopic theory through the values of these tensor coefficients. Given a
particular case these quantities could satisfy additional symmetries, which
would allow ad hoc approximations to achieve a concrete physical
characterization.

The $\phi $ field is dimensionless and thus the tensor $\lambda ^{\mu _{1}\mu
_{2}...\mu _{n}}$ must have a $(4-n)$ dimension, being typically of the order
of $\mu ^{4-n}$, where $\mu $ is the condensate scale. The low energy
requirement states that the typical scale $\epsilon $ of $\partial _{\nu}
\phi $ satisfies $\epsilon \ll $ $\mu $, which justifies us in expressing
the Lagrangian as an expansion on the derivatives of $\phi $ and in cutting
this expansion at order two. In the case where the system contains several
different energy scales a sensible expansion can involve different powers of
the space and time derivatives\cite{greiter}. This could be the case for
laboratory superfluids where the Fermi momentum and the fermion masses are
very different. We will focus on superfluids such as the one in a neutron
star, where these quantities are similar. For the description sketched above
to be valid during the whole evolution of the system, the gradient of $\phi $
must not only  be small at a given instant, but also at all times. From the
equation of motion of $\phi $, this implies that the coefficients $\lambda
^{\mu \nu }$ and $\lambda ^{\mu }$ must be sufficiently smooth.

The original theory is invariant under general coordinate transformations
and redefinitions of the field factorization. Once the background has been
established the effective theory admits only a restricted refactorization.
The effective second order Lagrangian will be invariant under the field
transformation $\phi \rightarrow \phi ^{\prime }=\phi -\Lambda $ , $\psi
\rightarrow \,\psi ^{\prime }=e^{i\,\Lambda (x)}\psi $ , if $\Lambda _{,\nu }
$ is of the same order as $\phi _{,\nu }$.

Inspired by the physical picture for the neutron star matter, where neutron
pairs have nonvanishing expectation values, we will assume that the U(1)
baryonic symmetry is broken to Z$_{2}$, the subgroup of transformations with
$\Lambda =0$ and $\Lambda =\pi $. This is similar to the superconductivity
effect, where there is a spontaneous breaking of the U(1) electromagnetic
gauge symmetry to Z$_{2}$. As $\phi $ parametrizes U(1)/Z$_{2}$, $\phi $ and
$\phi +\pi $ are taken to be equivalent.

The baryonic current is given by
\begin{equation}
j^{\mu }=\frac{\partial L}{\partial \phi _{,\alpha }}=\lambda ^{\mu \nu
}(x)\,\phi _{,\nu }+\lambda ^{\mu }(x)\;.  \label{jota}
\end{equation}
It contains two contributions. One is given by the Goldstone field, in terms
of the derivatives of $\phi $, and the other comes from the background
baryonic current given by $\lambda ^{\mu }$. The continuity equation for the
baryonic current, $\partial _{\mu }(\sqrt{g}j^{\mu })=0$, is equivalent to
the equation of motion of the Nambu-Goldstone field
\begin{equation}
D_{\mu }\left( \lambda ^{\mu \nu }(x)\,\phi _{,\nu }+\lambda ^{\mu
}(x)\right) =0\;.  \label{mov1}
\end{equation}

The velocity of charge transport can be defined by the relation $j^{\mu
}=n_{o}\,u^{\mu }$, where $n_{o}$ is interpreted as the charge density in
the fluid rest frame\cite{landau}. The superfluid state without vortices in
a flat space-time is characterized by the relation $\nabla \times {\bf
v}=0$, where ${\bf v}$ is the velocity of the fluid. In a curved
space-time, this relation is replaced by the one obtained from taking the
curl of the gradient in Eq. (\ref{jota})
\begin{equation}
\varepsilon ^{\alpha \beta \gamma \delta }D_{\delta }\left( \lambda
^{-1}\right) _{\gamma }^{\mu }\left( n_{o}u_{\mu }-\lambda _{\mu }\right)
=0\;.  \label{mov2}
\end{equation}
This last equation, together with Eq. (\ref{mov1}) defines the evolution of
a vortexless superfluid in a gravitational background.

The energy-momentum tensor is
\begin{equation}
T_{\nu }^{\mu }=\frac{1}{2}\left( \lambda ^{\mu \sigma }\,\phi _{,\sigma
}\phi _{,\nu }+\lambda _{\nu }^{\sigma }\,\phi _{,\sigma }\phi ^{,\mu
}\right) +\frac{1}{2}\left( \lambda ^{\mu }\phi _{,\nu }+\lambda _{\nu }\phi
^{,\mu }\right) -g_{\nu }^{\mu }\left( \frac{1}{2}\lambda ^{\sigma \tau
}\,\phi _{,\sigma }\phi _{,\tau }+\lambda ^{\sigma }\phi _{,\sigma }\right)
\;.  \label{em}
\end{equation}
The density of force acting on the fluid due to the interaction with the
background is $f^{\mu }=D_{\nu }T^{\nu \mu }=-D_{p}^{\mu }{\cal L}$, where
$D_{p}^{\mu}$ means the derivative with respect to the explicit dependence
of the Lagrangian in $x^{\mu }$. Thus
\begin{equation}
f^{\mu }=-\frac{1}{2}\left( \lambda ^{\sigma \tau }\right) ^{;\mu }\,\phi
_{,\sigma }\phi _{,\tau }-\left( \lambda ^{\sigma }\right) ^{;\mu }\;\phi
_{,\sigma }\;.
\end{equation}

We can define a global time by a time-like vector $\xi _{\mu }\sim \partial
_{t}$, with $\xi .\xi =1$, orthogonal to space-like surfaces that are set to
be equal time three-dimensional subspaces. The density of energy in such
equal time space-like subspaces is $\rho _{\xi }=\xi _{\mu }T_{\nu }^{\mu
}\xi ^{\nu }$:
\begin{equation}
\rho _{\xi }=\left( \xi _{\kappa }\,\lambda ^{\kappa \sigma }\xi
^{\tau }-\frac{1}{2}\lambda ^{\sigma \tau }\,\right) \phi _{,\sigma }\phi
_{,\tau }+\left( \left( \lambda \cdot \xi \right) \xi ^{\sigma }-\lambda
^{\sigma }\right) \phi _{,\sigma }\;,
\end{equation}
and the power exchanged with the background is given by the projection of
$f^{\mu }$ along $\xi _{\mu }$:
\begin{equation}
w=f^{\mu }\xi _{\mu }=-\frac{1}{2}D_{\xi }\left( \lambda ^{\sigma \tau
}\right) \,\;\;\phi _{,\sigma }\phi _{,\tau }-D_{\xi }\left( \lambda
^{\sigma }\right) \;\;\phi _{,\sigma }\;,
\end{equation}
which depends on the directional derivatives of $\lambda ^{\sigma \tau }$
and $\lambda ^{\sigma }$along $\xi $.

The covariant statement of the existence and stability of the ground state
can be expressed in terms of a minimum of the energy. If we want a local
extremum for $\phi _{,\sigma }=0$, it must be
\begin{equation}
\left( \lambda \cdot \xi \right) \xi _{\sigma }=\lambda _{\sigma }\;.
\end{equation}
This implies that $\xi_{\sigma}$ must be proportional to $\lambda_{\sigma}$,
the baryonic current for $\phi _{,\sigma }=0$. Thus the background
explicitly breaks the covariance of the theory. Furthermore, if the extremum
corresponds to a minimum, the matrix
\begin{equation}
M^{\sigma \eta }=\frac{\partial ^{2}\rho _{\xi }}{\partial \phi _{,\sigma
}\partial \phi _{,\eta }}=\left( \xi _{\rho }\,\xi ^{\eta }g_{\kappa
}^{\sigma }+\xi _{\rho }\,\xi ^{\sigma }g_{\kappa }^{\eta }-\,g_{\rho
}^{\eta }g_{\kappa }^{\sigma }\right) \lambda ^{\rho \kappa }
\end{equation}
must be positive definite. As stated above, if the Goldstone boson has a
true four dimensional dynamics, $\lambda _{\kappa }^{\rho }$ is regular,
and thus it has one time-like eigenvector and three space-like ones. If we
also require that the background be invariant under the time reversal
transformation $\xi _{\sigma }\rightarrow -\xi _{\sigma }$ , the $M^{\sigma
\eta }$ matrix becomes reducible into two non null submatrices: a
one-dimensional one in the subspace spanned by $\xi _{\sigma }$
\begin{equation}
\xi _{\sigma }M^{\sigma \eta }\xi _{\eta }=\xi _{\sigma }\lambda ^{\sigma
\eta }\xi _{\eta }\;,
\end{equation}
and a three-dimensional one in the space-like subspace orthogonal to $\xi
_{\sigma }$
\begin{equation}
\left( \xi ^{\rho }\xi _{\sigma }-g_{\sigma }^{\rho }\right) M^{\sigma \eta
}\left( \xi ^{\tau }\xi _{\eta }-g_{\eta }^{\tau }\right) =-\left( \xi
^{\rho }\xi _{\sigma }-g_{\sigma }^{\rho }\right) \lambda ^{\sigma \eta
}\left( \xi ^{\tau }\xi _{\eta }-g_{\eta }^{\tau }\right) \;.
\end{equation}
Therefore, in this case to have an energy minimum, the component $\xi
_{\sigma }\lambda ^{\sigma \eta }\xi _{\eta }$ of $\lambda ^{\sigma \eta }$
must be positive, whereas the matrix of the components in the orthogonal
subspace must be negative definite. Thus the minimum energy condition
guarantees dispersion relations with real propagation velocities for the
low-energy excitations. If the space-like submatrix is completely
degenerate, the propagation velocities are the same in all the spatial
directions and we have an isotropic background. Otherwise we have an
anisotropic one, with different propagation properties for the low energy
excitations according to the different spatial eigenvectors of $\lambda
^{\sigma \eta }$.

The spacial components $u_{i}$ of the velocity are proportional to $\frac{1}
{n_{o}}\lambda_{i}^{\mu }\phi _{,\mu }$ and this implies that the low
energy expansion is also a small velocity expansion $u_{i}\ll 1$.

We can implement a hydrodynamical description of the fluid by writing the
energy momentum tensor in terms of the current. It reads
\begin{eqnarray}
T_{\nu }^{\mu }=&&\frac{n_{o}}{2}\left( u^{\mu }\left( \lambda ^{-1}\right)
_{\rho \nu }\left( n_{o}u^{\rho }-\lambda ^{\rho }\right) +u_{\nu }\left(
\lambda ^{-1}\right) _{\rho }^{\mu }\left( n_{o}u^{\rho }-\lambda ^{\rho
}\right) \right)  \nonumber\\
&&-g_{\nu }^{\mu }\frac{1}{2}\left( \lambda ^{-1}\right)
_{\rho \tau }\left( n_{o}^{2}u^{\tau }u^{\rho }-\lambda ^{\tau }\lambda
^{\rho }\right) \;.
\end{eqnarray}
This tensor only contains the terms corresponding to the dynamics of the
excitations with respect to the background. To have the energy momentum
tensor of the fluid as a whole, it is necessary to add the contribution of
the rest of the fluid. Assuming that the fluid is an isotropic perfect one
up to first order in the velocities, we have
\begin{equation}
\mu =\frac{(\rho +p)}{n_{o}}=\frac{n_{o}}{2}\left( \lambda ^{-1}\right)
_{i}^{i}\;,
\end{equation}
where $\mu $ is (for zero temperature) the chemical potential of the global
conserved number, $\rho $ is the energy density and $p$ is the pressure of
the fluid.

\section{Vortices and the generalized Magnus force}

We have not considered the existence of vortices up to now, but they will be
present if the fluid has some angular moment. In general, the superfluid
cannot rotate as a whole, and therefore if it has a non null angular
momentum it must be supported by vortices. It is difficult to analyze the
vortex-fluid interaction in the Goldstone field representation for the
superfluid, because the vortex imposes non trivial boundary conditions to
the scalar field. In the large distance limit where the vortex size could be
considered negligible these conditions manifest themselves as a field
nondifferentiability that leads to a noncommutativity of its partial
derivatives. A more suitable approach could be developed relying on the dual
description of the scalar field in terms of a Kalb-Ramond one, an
antisymmetric tensor of second order $B_{\mu \nu }$\cite{davis,libro}. Both
massless theories, a k-form field and a (n-k-2)-form field, where n is the
space-time dimension, are equivalent not only at a classical level, but also
at the quantum one. This point will be discussed in detail elsewhere\cite
{kaul}. We will use a generalized version of this duality here.

The dual variables are defined by expressing the Noether conserved baryonic
current as a topologically conserved current. This is achieved with the
identification

\begin{equation}
J^{\lambda }=\frac{1}{6}E^{\mu \nu \rho \lambda }H_{\mu \nu \rho }\;\,,
\label{corr}
\end{equation}
where $H_{\mu \nu \sigma }=\partial _{\mu }B_{\nu \sigma }+\partial _{\nu
}B_{\sigma \mu }+\partial _{\sigma }B_{\mu \nu }$ and $E$ is the Hodge
tensor with covariant and contravariant components $E_{\mu \nu \rho \lambda
}=\sqrt{-g}\varepsilon _{\mu \nu \rho \lambda }$ and $E^{\mu \nu \rho
\lambda }=-\frac{1}{\sqrt{-g}}\varepsilon ^{\mu \nu \rho \lambda }$
respectively. The Levi-Civita symbols denote the index permutation
signature. The completely antisymmetric tensor $E$ commutes with the
covariant derivatives. The inverse relation to (\ref{corr}) is $H_{\mu \nu
\rho }=-E_{\mu \nu \rho \lambda }J^{\lambda }$.

The divergence of the current $J^{\lambda }$ is proportional to the exterior
derivative $d\wedge H$, and nullifies because $H=d\wedge B$. In components
it reads
\begin{equation}
\varepsilon ^{\mu \nu \rho \lambda }\partial _{\lambda }H_{\mu \nu \rho
}=0\;\,.
\end{equation}
In this way the equation of motion for the field $\phi$, which is
equivalent to the baryonic current conservation, is satisfied automatically
in a topological sense in the Kalb-Ramond field scheme.

In a similar way, the topological property of the scalar field $d\wedge(d
\wedge \phi )=0$ becomes the equations of motion for the Kalb-Ramond
field $B_{\mu \nu }$, by means of the duality transformations. Duality
states a correspondence between the physical solutions of the two theories.
This correspondence is realized as a canonical transformation at the level
of Dirac brackets, which define the quantum theory\cite{kaul}.

Both descriptions are linked by Eq. (\ref{corr}), which is a relation
between the field derivatives
\begin{equation}
\phi _{,\kappa }=\left( \frac{1}{6}E^{\mu \nu \rho \sigma }H_{\mu \nu \rho
}-\lambda ^{\sigma }\right) \lambda _{\sigma \kappa }^{-1}\,\;,
\label{relat}
\end{equation}
where the tensor $\lambda _{\sigma \kappa }^{-1}$ is the inverse of $\lambda
^{\sigma \kappa }$. This relation is a generalization of the one used in
Ref. \cite{davis}, where only the particular case $\lambda _{\sigma \kappa
}\propto g_{\sigma \kappa }$ is considered. The curl of $\phi _{,\kappa }$
is null, and this implies that the curl of the right hand side must also be
null. This gives the equations of motion for the $B$ field:
\begin{equation}
E^{\alpha \beta \gamma \kappa }D_{\gamma }\left( \frac{1}{6}E^{\mu \nu \rho
\sigma }H_{\mu \nu \rho }\lambda _{\sigma \kappa }^{-1}-\lambda _{\sigma
\kappa }^{-1}\lambda ^{\sigma }\right) =0\;\,.
\end{equation}

The corresponding action that leads to this equation can be written as
\begin{equation}
{\cal S}_{B}=\frac{\kappa }{3}\int d^{4}x\sqrt{-g}\;H_{\alpha \beta \gamma
}E^{\alpha \beta \gamma \kappa }\left( \frac{1}{12}E^{\mu \nu \rho \sigma
}H_{\mu \nu \rho }\lambda _{\sigma \kappa }^{-1}-\lambda _{\sigma \kappa
}^{-1}\lambda ^{\sigma }\right) \,.
\end{equation}
Equating the energy-momentum tensor from this action with the one for the
scalar field, Eq. (\ref{em}), using the relation (\ref{relat}), we get
$\kappa =-\frac{1}{2}$.

The 2-form Kalb-Ramond field naturally couples with the two dimensional
manifold generated by the world sheet of the vortex moving along time. This
formalism is advantageous because it allows us to describe the fluid-vortex
interaction with an explicit term in the action, without using any extra
boundary conditions. Using the antisymmetric tensor $J^{\mu \nu }=\eta \int
d\sigma ^{\mu \nu }\delta ^{4}\left( x-y(\sigma ,\tau )\right) $, which
describes the surface that sweeps the vortex in space time, with the
coordinates $\left( \sigma ,\tau \right) $ parametrizing the world sheet,
the fluid-vortex interaction term is
\begin{equation}
{\cal L}=B_{\mu \nu }J^{\mu \nu }\;,
\end{equation}
a metric independent term because it corresponds to the surface integral of
a two form. It is possible to think of different interaction terms that
depend on the metric or that have field derivatives, but for dimensional
reasons those terms should be suppressed. They must contain additional
powers of the superfluid coherence length over the vortex curvature scale or
the space curvature, and can be neglected in this context.

Taking into account the interaction with the vortex, the equation of motion
is written
\begin{equation}
\frac{1}{2}E^{\alpha \beta \gamma \kappa }D_{\gamma }\left( \frac{1}{6}
E^{\mu \nu \rho \sigma }H_{\mu \nu \rho }\lambda _{\sigma \kappa
}^{-1}-\lambda _{\sigma \kappa }^{-1}\lambda ^{\sigma }\right) =J^{\alpha
\beta }\;.  \label{fuerza}
\end{equation}
This equation describes the possible singularities of the Goldstone field
since the first term translates as the curl of the gradient of $\phi $.
Using the Stokes theorem, the flux of $J^{\alpha \beta }$ is related with
the circulation $\oint \bigtriangledown \phi \,ds=k\pi $ around the vortex
lines, where $k$ is an integer. This condition is imposed in the Kalb-Ramond
field scheme by $\eta =k\frac{\pi }{4}$ , where $k$ is the vortex
topological number.

Equation (\ref{fuerza}) provides the general relation between current
and vorticity
\begin{equation}
\frac{1}{2}E^{\alpha \beta \gamma \kappa }D_{\gamma }\left( \lambda _{\sigma
\kappa }^{-1}\left( J^{\sigma }-\lambda ^{\sigma }\right) \right) =J^{\alpha
\beta }\;\,,  \label{jv}
\end{equation}
and allows us to infer the vortex distribution from the current. Using the
Stokes theorem the circulation $\frac{1}{2}\oint dx^{\kappa }\lambda
_{\sigma \kappa }^{-1}\left( J^{\sigma }-\lambda ^{\sigma }\right) $ in the
perimeter is equal to the flux $\int dx^{\mu }dx^{\nu }E_{\mu \nu \alpha
\beta }J^{\alpha \beta }$ over the enclosed surface, and we can write
\begin{equation}
\oint \lambda _{\sigma \kappa }^{-1}\left( J^{\sigma }-\lambda ^{\sigma
}\right) dx^{\kappa }=\pi N_{v}\,,  \label{circ}
\end{equation}
where the second term is the flux of $J^{\alpha \beta }$, and represents
the total number of vortices enclosed in the integration path.

The presence of a vorticity gives an additional term to the force acting on
the fluid. The force is given by
\begin{equation}
\tilde{f}^{\nu }=D_{\mu }T^{\mu \nu }=-J_{\mu \sigma }H^{\mu \sigma \nu
}-J^{\kappa }D^{\nu }\left( \lambda _{\sigma \kappa }^{-1}\lambda ^{\sigma
}\right) +\frac{1}{2}J^{\kappa }J\,^{\sigma }\,D^{\nu }\left( \lambda
_{\sigma \kappa }^{-1}\right) \;,  \label{fff}
\end{equation}
where the first term on the right hand side comes from the new term in the
equation of motion for $H$, and the Bianchi identities have been used. The
last terms are only relevant in regions where the background has a large
gradient, with a scale of the order of the coherence length, as happens for
the dynamics of the vortices in the crust of a neutron star where the
vortex pins.

We will now calculate the force acting on a single vortex. In this case the
solution of (\ref{fuerza}) could be written as a background solution, whose
source are the other vortices, $B_{o}^{\mu \nu }$, plus a particular
solution with the vortex as a source, $B_{v}^{\mu \nu }$, i.e. $H^{\mu
\sigma \nu }=H_{o}^{\mu \sigma \nu }+H_{v}^{\mu \sigma \nu }$. The term
$H_{o}^{\mu\sigma\nu }$ can be written using relation (\ref{corr}),
from the average current
\begin{equation}
H_{o}^{\mu \sigma \nu }=-E^{\mu \sigma \nu \rho }J_{\rho }\;\,.
\end{equation}
We are not taking into account the dynamics of the vortex, which could be
responsible for tension effects but is not given by the effective theory.
These effects are related to interaction terms between $J_{\mu \sigma }$ and
$H_{v}^{\mu \sigma \nu }$, which are not relevant if the curvature radius of
the vortex is much greater than the coherence length. Considering also
that the background potentials have a slow variation along the vortex size,
the force on the vortex per unit length becomes
\begin{equation}
f^{\nu }=-\tilde{f}^{\nu }\simeq -J_{\mu \sigma }E^{\mu \sigma \nu \rho
}J_{\rho }\;.  \label{fv}
\end{equation}
In general for a vortex in equilibrium the Magnus force compensates the
total external forces acting on it. Considering only the interactions here
discussed, the equilibrium is specified by the null Magnus force condition,
which can be put in terms of the current as
\begin{equation}
J^{\kappa }\left[ D_{\delta }\left( \lambda _{\sigma \kappa }^{-1}\left(
J^{\sigma }-\lambda ^{\sigma }\right) \right) -D_{\kappa }\left( \lambda
_{\sigma \delta }^{-1}\left( J^{\sigma }-\lambda ^{\sigma }\right) \right)
\right] =0\;\,.  \label{nf}
\end{equation}

The vortex can be locally characterized by its tangent vector $m^{\nu }$ and
its quadrivelocity $v^{\nu }$. Thus we have
\begin{equation}
d\sigma ^{\mu \nu }=\left( v^{\mu }m^{\nu }-m^{\mu }v^{\nu }\right) d\sigma
d\tau \;,
\end{equation}
where $\sigma $ is defined such that $m^{\nu }m_{\nu }=1$ and $\tau $ is the
proper time on the vortex, so $v^{\nu }v_{\nu }=-1$. Using $J^{\mu
}=n_{0}u^{\mu }$, where $u^{\mu }$ is the fluid quadrivelocity, and
integrating over a transversal cut, the force per unit length finally
results
\begin{equation}
f^{\mu }=-\frac{\pi }{2}n_{o}\sqrt{-g}g^{\mu \rho }\epsilon _{\rho \nu
\sigma \tau }u^{\nu }v^{\sigma }m^{\tau }\;.   \label{magnus}
\end{equation}

This is the expression for the generalized Magnus force. Introducing $\delta
u^{\mu }=v^{\mu }-u^{\mu }$ and considering that to first order in the
velocities $\delta u^{0}=0$, the 3-d force acting on a vortex with $m^{\mu
}=(0,{\bf {m})}$ is
\begin{equation}
{\bf f}=\frac{\pi }{2}n_{0}\sqrt{-g}\sqrt{-g^{00}}\left[ {\bf G}{\cdot }
\left( {\bf \delta u}{\times }{\bf m}\right) \right] {\bf \;,}
\end{equation}
where ${\bf G=(}g^{ij})$, and ${\bf f=}\left( f^{i}\right) $, and the power
dissipated by the interaction of the vortex with the fluid is
\begin{equation}
f^{0}=\frac{\pi }{2}n_{0}\sqrt{-g}\sqrt{-g^{00}}\left( \sqrt{-g^{00}}{\bf
v.u\times m}+2{\bf g}.{\bf \delta u}{\times }{\bf m}\right) \;.
\end{equation}
In flat space-time this force reduces to the ordinary Magnus force ${\bf f}=
\frac{\pi }{2}n_{0}\left( {\bf \delta u}{\times }{\bf m}\right)$.

The Magnus force implies that the equilibrium is reached when the vortex
moves at the same velocity as the medium, or when the vortex is aligned with
the relative velocity between the vortex and the background.

\section{Examples}

In this section we present two examples of superfluids in a gravitational
field. The first is a superfluid in a laboratory on Earth. It is simple, but
its relevance is only academic because the interesting effects are extremely
weak. On the other hand, the second example is of great phenomenological
interest, because it corresponds to a superfluid in a neutron star, which is
one of its main components.

In the construction of the effective Lagrangian theory we include all the
terms consistent with the symmetries of the problem and relevant for the
energy scale that matters. In general there is more than one Goldstone boson
that arise from the symmetry breaking. In the following examples we consider
only one boson. This is true for He$^{4}$ superfluid but not for He$^{3}$ or
the neutron star case, where part of the superfluid can be in the spin two
$^{3}P_{2}$ state. In spite of this the baryonic phase can be factorized
from the spin degrees of freedom. Specifically, for the neutron star the
effect of the interaction of the baryonic Goldstone boson with these spin
degrees is negligible\cite{pines}. This is easy to see by considering that
the ratio between the spin angular moment and the orbital angular moment for
a single neutron is $\frac{\hbar }{\Omega R^{2}m}\ll 1$, indicating that the
spin can not significantly alter the global dynamics of the baryonic
current.

\subsection{Superfluid in a laboratory on Earth}

In this case we have a weak gravitational field without a noticeable
contribution from the fluid itself, with a time independent metric that can
be written:
\begin{eqnarray}
g^{00} &=&1-2U\;, \\
g^{ij} &=&-\left( 1+2U\right) \delta ^{ij}\;, \\
g^{0i} &=&h^{i}\;,
\end{eqnarray}
where $U=g\,z$ is the Newtonian potential and $\vec{h}$ gets contributions
from the dragging and inertial effects due to the Earth rotation. The
existence of an energy extremum for a null $\phi $ gradient implies that
$\lambda_{\mu}$ is proportional to the time derivative Killing vector, with
$\lambda _{0}=n(z)\;$and$\;\lambda _{i}=0$. Here the minimum energy
condition is equivalent to $\lambda^{00}>0$ and to the matrix $\lambda^{ij}$
negative definite. In general both the gravitational dragging and the fluid
velocities are small, so that the second order tensor $\lambda ^{\mu \nu }$
can be considered independent of $\vec{h}$. Taking into account the
rotational symmetry on the vertical $\hat{z}$ axis, the traslational
symmetry in the horizontal $(x,y)$ plane, and the invariance under temporal
inversion, the non null components are
\begin{equation}
\lambda _{0}^{0}=a(z)\;,\;\lambda _{x}^{x}=\lambda _{y}^{y}=\lambda
_{z}^{z}=b(z)\;.
\end{equation}
Here $a$ and $b$ are scalars that depend only on the fluid conditions. We
take $\lambda _{x}^{x}=\lambda _{y}^{y}=\lambda _{z}^{z}$ because the
gravitational field scale is much longer than the microphysics scale that
determines the superfluid state, so that any possible microscopic anisotropy
induced by the gravitational field should be negligible. The current
can be written as
\begin{equation}
j_{0}=a\dot{\phi}+n\,\,\,\,\,\,\,\,\,\,\,\,\,,\,\,\,\,\,\,\,\,\,\,\,j_{i}=b
\partial _{i}\phi \;.
\end{equation}
Let us consider the stationary case. The charge density in the fluid rest
frame $n_{o}$ is equal to $n$ at first order in the velocity. The
tridimensional velocity is given by the contravariant components
\begin{equation}
\vec{v}=\frac{1}{n_{o}}(j^{i})=\vec{h}-\frac{b}{n_{o}}\left( 1+2U\right)
\vec{\nabla}\phi \;.  \label{grad}
\end{equation}
This implies $\vec{\nabla}\times \,\left( \frac{n_{o}\left( 1-2U\right) }{b}
(\vec{v}-\vec{h})\right) =0$, a generalization of the usual superfluid
relation $\vec{\nabla}\times \vec{v}=0$. This equation is analogous to the
superconductor equation $\vec{\nabla}\times
\,(\vec{v}-\frac{e}{m}\vec{A})=0$, but there is a subtle detail in this
analogy. In our case the origin of the velocity drift is the metric and not
the connection, as in the superconductor case. Thus for a superfluid the
velocity field in the fundamental state will not be null, but equal to
$\vec{h}$. The term with the gradient of the field in Eq. (\ref{grad}) is non
null when there are vortices contributing to the fluid rotation. As can be
deduced from this equation the contribution of $\vec{h}$ generally
diminishes the number of vortices with respect to a flat space situation.

When $\vec{h}$ has an inertial origin, due to choosing a rotating reference
frame, we have $\vec{\nabla}\times \vec{h}=-\vec{w}$, the angular velocity
vector. This tells us that the superfluid is in fact non rotating. Inertial
effects on superfluids have been used in the construction of a precision
gyrometer\cite{avenel}. In the case of a superconductor, inertial effects
lead to a compensating London field\cite{london}, which is also induced by
the gravitational dragging\cite{deWitt}.

The effective mass of the boson condensate $m^{*}$ can be identified from
Eq. (\ref{grad}) at null gravitational field, as the coefficient of
$\vec{\nabla}\phi .$ It results $m^{*}=n_{o}/b$. The Lagrangian density is
\begin{equation}
L=\left( 1-2U\right) \left( a\dot{\phi}^{2}-b\left( \vec{\nabla}\phi \right)
^{2}+n\dot{\phi}\right) +\left( 2a\dot{\phi}+n\right) \left( \vec{h}\cdot
\vec{\nabla}\phi \right) -4bU\left( \vec{\nabla}\phi \right) ^{2}\;.
\end{equation}
The limit of zero gravitational field allows us to clarify the physical
meaning of $n$, $a$ and $b$, by comparing with the low energy BCS
Lagrangian\cite{schakel}. Thus $n=k_{F}^{3}/3\pi ^{2}$, with $k_{F}$ the
Fermi wave number, is the charge density. We can also relate $a$ and $b$ with
the Fermi wave number and the fermion mass $m$, $a=mk_{F}/2\pi ^{2}$ and
$b=k_{F}^{3} /6\pi ^{2}m$, and thus the sound velocity is given by
$\sqrt{\frac{b}{a}}= \frac{k_{F}}{m\sqrt{3}}=\upsilon _{F}/\sqrt{3}$, where
$\upsilon _{F}$ is the Fermi velocity. The relation between the fermion mass
and the effective pair mass is $m^{*}=2m$. Finally, the factor $(1-2U)$
corresponds to a redshift effect.

If there are vortices, the expression (\ref{magnus}) gives the Magnus force
acting upon them. In this case we have the usual flat space-time force only
affected by a redshift factor.

\subsection{Superfluid neutron star}

Here we develop with some detail the example of the superfluid component in
a neutron star. In the case of a neutron star at rest we can assume that the
metric satisfies a spherical symmetry
\begin{equation}
\left( ds_{o}\right) ^{2}=e^{2\Phi }dt^{2}-e^{2\Lambda }dr^{2}-r^{2}(d\theta
^{2}+\sin {}^{2}\theta \ d\varphi ^{2})\;.  \label{m1}
\end{equation}
The potentials $\Phi (r)$ and $\Lambda (r)$ that characterize this metric
are given by the Oppenheimer-Volkoff equations for cold stars\cite{misner}.

In the case of neutron stars the friction is high enough to drive the star
to equilibrium in relatively short times. This allows us to assume a
stationary neutron star, which rotates at a small angular velocity $\Omega$.
This magnitude represents the angular velocity measured by an observer at
rest with the fluid, and is related with the fluid velocity by $\Omega
=u^{\varphi }/u^{t}$, such that
\begin{equation}
j^{\mu }=n_{o}u^{0}(1,0,0,\Omega )\;.
\end{equation}

We can make an expansion of the rotating star metric in the perturbation
with respect to the rest star metric and the angular velocity $\Omega $. The
source for this perturbation is the superfluid energy momentum tensor. It
can be considered a small quantity for ordinary pulsars because the
gravitational acceleration, approximately $\frac{GM}{R^{2}}$, is much
greater than the centrifugal one, $\Omega ^{2}R$. As an example, for the
rapidly rotating pulsar Crab the quotient of these accelerations is
approximately $50$, implying small deformations. Hence we can keep the
expansion up to the first order in the fluid angular velocity\cite{hartle}
\begin{equation}
ds^{2}=\left( ds_{o}\right) ^{2}-2r^{2}\sin {}^{2}\theta \ \omega \,d\varphi
\,dt\;.
\end{equation}
The gradients $\phi _{,\nu }$ can also be considered first order quantities
because they can be chosen to be zero for the star at rest.

In the construction of the effective Lagrangian theory we will consider all
the terms that satisfy the symmetries of the problem and are relevant for
the energy scale that matters. We can identify several symmetries: the
baryonic global symmetry, the covariance under general coordinate
transformations, the spherical symmetry, the temporal translation and the
temporal reflection symmetry of the star at rest.

The time translation and rotational invariance makes $\lambda _{\mu
}=f(r)\,\chi _{\mu }$, where $\chi _{\mu }=(e^{2\Phi },0,0,0)$ is the time
Killing vector. The covariant components of this Killing vector are non
dynamical, contrary to the contravariant components, which depend on the
metric perturbations.

Time translation and spherical symmetry impose important restrictions to
$\lambda ^{\mu \sigma }$. We must have
\begin{equation}
\lambda _{0}^{0}=a(r)\;\;,\;\;\lambda _{r}^{r}=b(r)\;\;,\;\;\lambda _{\theta
}^{\theta }=c(r)\;\;,\;\;\lambda _{\varphi }^{\varphi }=c(r)\;\,,
\end{equation}
where $a$, $b$ and $c$ are scalar quantities, because they are the
eigenvalues of a mixed tensor. The stability of the ground state requires
that $a,$ $b,$ and $c$ must be positive. The metric perturbations do not
affect the quadratic term in $\phi _{,\sigma }$ because this is already of
second order.

The interpretation of $b$ and $c$ comes from the equations
\begin{eqnarray}
u_{r} &=&\frac{b(r)}{n_{0}(r)}\partial _{r}\phi =\frac{1}{m_{r}^{*}(r)}
\partial _{r}\phi \;\,, \\
u_{\theta } &=&\frac{c(r)}{n_{0}(r)}\partial _{\theta }\phi =\frac{1}
{m_{\theta }^{*}(r)}\partial _{\theta }\phi \,\;, \\
u_{\varphi } &=&\frac{c(r)}{n_{0}(r)}\partial _{\varphi }\phi =\frac{1}
{m_{\varphi }^{*}(r)}\partial _{\varphi }\phi \,\;,
\end{eqnarray}
where $m_{r}^{*}=n_{0}/b$, $m_{\theta }^{*}=m_{\varphi }^{*}=n_{0}/c$ are
scalar quantities of dimension one that depend only on the fluid conditions,
and because of this are functions of the radius. By using the equivalence
principle, these parameters can be interpreted as the effective mass of the
bosonic quasiparticles in the different directions. This is the double of the
fermionic particle effective mass, which can be calculated from the
particular model as the ratio between the Fermi momentum and the Fermi
velocity.

The quantities that determine the fluid conditions are related to each other
independently of the gravitational field. This is because this field can be
considered constant along the microscopic nuclear interaction range that
determines the plasma dynamics. Because of this, and as long as the Fermi
surface of the fluid has no preferred direction, we can take $m_{r}^{*}=
m_{\theta }^{*}=m_{\varphi }^{*}=m^{*}$. For a neutron star $m^{*}$ includes
plasma effects and has values of the order of twice the neutron
mass\cite{wiringa}.

For the star at rest we can identify $j_{0}=$\thinspace \thinspace $\lambda
_{0}=n_{o}e^{\Phi }$. Up to first order in $\omega $, for a rotating star
$u_{0}$ is the same as for a star at rest, and thus this\thinspace equation
is also valid for a rotating star up to first order in the angular velocity.

The relation between the angular velocity $\Omega $ and the metric
coefficient $\omega $ in absence of vortices is given by Eq.(\ref{mov2}).
The solution of this equation is $\Omega =\omega (r,\theta )$ , i.e. the
angular velocity of the fluid must be the same as the angular velocity that
takes an object that falls free from infinity to the point $(r,\theta )$,
and corresponds to the angular velocity of local inertial frames with
respect to the fixed stars.

Furthermore, in the case we are discussing here, the star that produces the
gravitational field is formed by the rotating fluid. Thus $\omega $ must
satisfy the Einstein equation for the $R_{\varphi t}$ Ricci tensor
component, with the energy momentum component $T_{\varphi t}$ of the fluid
as the source. The solution for the perturbed Einstein equation is a $\theta
$-independent $\omega $ that satisfies
\begin{eqnarray}
\omega ,_{rr} &+&(\frac{4}{r}-\Lambda ^{\prime }-\Phi ^{\prime }\,)\,\omega
,_{r}+\frac{2}{r}\,(\frac{1}{r}+\Phi ^{\prime }-\Lambda ^{\prime }-\frac{1}
{r}e^{2\Lambda })\,\omega   \nonumber \\
&=&8\pi e^{2\Lambda }((\rho +3p)\,\omega \,-2\,(\rho +p)\,\Omega \,\,)\;\,.
\label{e}
\end{eqnarray}
For nonsingular gravitational potentials $\Lambda $ and $\Phi $ the only
solution for $\Omega =\omega $ with regular geometry is $\omega =0$\cite
{casini}. That means that there is no rotating star solutions without
vortices, i.e. a rotating superfluid star necessarily contains vortices, in
which case $\left( \Omega -\omega \right) $ is non null. The background acts
on these vortices with a force given by Eq. (\ref{fv}). If this is the only
force upon the vortex, from it we can obtain the vortex profile in the
stationary state. In this case the vortex adapts to the background in such a
way that this force becomes null. This configuration is given by Eq. (\ref
{nf}), and in the case we are discussing here it is consistent with
$j^{0}=\lambda ^{0}=n_{o}e^{-\Phi }$ and $j^{\varphi }-\lambda ^{\varphi
}=n_{o}e^{-\Phi }\left( \Omega -\omega\right)$. Furthermore, Eq. (\ref{jv})
gives the relation between the current $j^{\mu }$ and the vorticity
$J^{\mu\nu}$, and allows us to compute the distribution and orientation of
the vortices. According to Eq. (\ref{circ}), in this case we have
\begin{equation}
\Omega -\omega (r,\theta )=\frac{N_{v}(r,\theta )}{m^{*}e^{-\Phi (r)}}
\frac{1}{r^{2}\sin ^{2}\theta }\;,  \label{vort}
\end{equation}
where $N_{v}$ is the number of vortices within a closed circular path that
passes by the point $(r,\theta )$ and is perpendicular to the $\hat{z}$
axis, $m^{*}$ is the effective mass, and $e^{-\Phi }$ gives the red shift of
the effective mass due to the gravitational field.

For a rotating star at the minimum energy configuration $\Omega $ is
constant, and thus Eq. (\ref {vort}) together with Eq. (\ref{e}) allows us
to completely determine the vortex profile. The angle $\beta $ between the
axis of the star and the direction of the vortex lines at a given point
$(r,\theta )$ can be deduced from Eq. (\ref{vort}), using the cylindrical
symmetry of the system. It reads \begin{equation} \sin \beta
=\frac{1}{2}\frac{\kappa (r)\;\sin (2\theta )}{\left( 1+\kappa (r)(\kappa
(r)-2)\;\sin {}^{2}(\theta )\right) ^{1/2}}\;\,, \end{equation}
$\,$where $\kappa (r)=\frac{r}{2}\left( \frac{1}{\Omega -\omega }\frac
{d\omega}{dr}-\frac{1}{m*}\frac{dm*}{dr}+\frac{d\Phi}{dr}\right)$. The
first term in $\kappa $ corresponds to a purely geometrical contribution due
to the Lense-Thirring effect, whereas the two last terms depend on the
microscopic structure of the fluid and the star as a whole.

In particular, if we take $r$ equal to the radius $R$ of the star and
$\theta =\frac{\pi }{2}$, we have $\omega =\frac{2GJ}{R^{3}}$, where $J$ is
the angular momentum of the star, and thus the total number of vortices
contained by the star is
\begin{equation}
N=m^{*}e^{-\Phi \left( R\right) }\left( \Omega R^{2}-\frac{2GJ}{R}\right) \;.
\end{equation}
A noticeable effect of the gravitational background is a decrease of the
vortex density with respect to a similar situation in flat space-time. In
the case of a typical neutron star this decrease is of the order of 15\%.

\section{Final remarks}

We have considered the superfluidity phenomena in presence of a
gravitational background. Our starting point is the spontaneous symmetry
breaking of a U(1) symmetry. This field-theoretical approach allows us to
develop from first principles a fully covariant formalism. Within this
framework we study general aspects of the dynamics of the superfluid in a
gravitational field. We analyze the force acting on a vortex, which is a
generalization of the well known Magnus force, and in particular we find the
profile of a vortex in equilibrium with the condensate. This approach makes
contact with the microscopic theory which describes the details of the
superfluidity phenomena in two points. One is a very basic and general one,
with a strong theoretical support, which is the spontaneous symmetry
breaking mechanism. The other point of contact is the specific description
of the background through the tensors $\lambda ^{\mu \nu }$ and $\lambda
^{\mu }$, which in general must satisfy some symmetry requirements, but
their details depend on the microscopic physics of the system. In some
relatively simple cases it is possible to construct a rather closed
description, dependent only on a few phenomenological parameters, making use
of the known symmetries.

Our results are exemplified with two systems. One of them is a superfluid in
a terrestrial laboratory, where we have a weak gravitational field. The main
result here is the formal analogy with a superconductor, with the
gravimagnetic field $\vec{h}$ playing the role of the vector potential
$\vec{A}$, and the Newtonian potential $U$ the role of the Coulomb
potential. This analogy is not complete, because the geometrical origin of
these fields is different. In the gravitational case $\vec{h}$ is introduced
by the metric, whereas the $\vec{A}$ field is due to the connection. One
consequence of this difference is that in the electromagnetic case we have a
Meissner effect, while in the gravitational case we have an anti-Meissner
one\cite {casini}.

The other example considers a more interesting system, which is the
superfluid in a neutron star. In this case we have a strong gravitational
field, which makes a fully covariant treatment unavoidable. Here we
construct a description of the system based on the symmetries of the star,
and this is enough to determine the shape and distribution of the vortices,
assuming that there are no external forces. This should be the case in the
star core where there are no pinning forces. Other forces, such as the
magnetic ones, cannot appreciably alter the vortex distribution because
their energy is much smaller than the rotational energy. The generalized
Magnus force here analyzed could be very relevant for a detailed study of
the transient during the pulsar glitches, if we have an adequate model for
the pinning forces.

\section*{Acknowledgments}

This work was realized with a partial support from the Consejo Nacional de
Investigaciones Cient\'{\i }ficas y T\'{e}cnicas (CONICET), Argentina.

\end{document}